\documentclass[preprint,12pt]{elsarticle}
\usepackage{amssymb}
\usepackage{amsmath}

\usepackage{graphicx} 
\usepackage{geometry}
\usepackage[hidelinks]{hyperref}
\usepackage[utf8]{inputenc}
\usepackage{subcaption}
\usepackage{amsmath}
\usepackage{bm}
\usepackage{tikz}
\usepackage{url}
\usetikzlibrary{calc,patterns,positioning,angles,quotes}

\journal{Nuclear Instruments and Methods in Physics Research Section A} 

\begin{document}

\begin{frontmatter}

\title{A parametrised simulation code to simulate the response of ground detector arrays to Extensive Air Showers above 100 TeV}

\author[UniTo,INFN,]{Andrea Chiavassa}
\author[UniTo,INFN]{Andrea Negro}
\affiliation[UniTo]{organization={University of Turin, Department of Physics},
	addressline={Via Pietro Giuria 1}, 
	city={Torino},
	postcode={10125}, 
	state={},
	country={Italy}}
\affiliation[INFN]{organization={INFN-Turin},
	addressline={Via Pietro Giuria 1}, 
	city={Torino},
	postcode={10125}, 
	state={},
	country={Italy}}

\begin{abstract} 
Surface arrays detecting the secondary particles of Extensive Air Showers (EAS) are one of the techniques used to detect gamma-rays from high-energy sources. To evaluate the reconstruction performance, gamma/hadron separation, and sensitivity of such experiments, many simulations of gamma rays and protons are required. The usual approach is divided into two steps. First, the CORSIKA software is used to generate EAS at the observation level. Then, a GEANT4-based full simulation code is used to simulate the array response. For very large fill factors, above 30-100 TeV, the computational resources required become too demanding.  In this work, the authors present an alternative way to evaluate the performance-related quantities of a simulated array configuration through the parametrisation of the detector response. We will present the implementation of such software, the validation checks we performed, and the application of this tool to a representative future ground-based water-Cherenkov detector array.
\end{abstract}




\end{frontmatter}

\section{Introduction}
The recent results of the HAWC \cite{hawc} and LHAASO \cite{lhaaso} experiments have started the study of galactic gamma-ray sources in the range from 100 TeV to a few PeV. Such sources have been called PeVatrons and are among the brightest gamma ray accelerators in our Galaxy. While familiarity with these sources is not required to follow this work, additional background can be found in the following review articles \cite{PeVatronHAWC,PeVatronsTA,PeVatrons,UHEAstronomy,huntPeVatrons}.

Both LHAASO and HAWC experiments are located in the northern hemisphere, but the first LHAASO catalogue \cite{LHAASOCatalogue} indicates that such sources are concentrated in the galactic plane. Therefore, we can expect that Ultra High Energy (UHE) sources are present in the galactic centre. The Southern Wide-field Gamma-ray Observatory (SWGO) \cite{swgo2019wp,swgo} will be built in the southern hemisphere and will be able to study such area of the sky in a better zenith range. The galactic centre is visible from the northern hemisphere only at high zenith. 

The SWGO collaboration aims to build a large Field-of-View (FoV) ground array with high duty cycle ($\sim100\%$), studying primary gamma rays from 100 GeV to a few PeV. It will be located at high altitude (4770 m) in Pampa la Bola in the Atacama astronomical park\footnote{Webpage: \url{https://astronomicalpark.cl/}} in Chile ($-22.9^\circ$ latitude and $-67.7^\circ$ longitude). It will detect the secondary particles of Extensive Air Showers (EAS) at the observation level using an array of Water Cherenkov Detectors (WCDs) equipped with one or more PMTs.

An experiment aiming to study PeVatrons requires a high effective area and a strong gamma/hadron (g/h) separation power ($10^4-10^5$). For this purpose, the collaboration is exploring analysis methods that exploit both the morphology of the events and the muon content of the showers, using numerical variables \cite{PINC_LDFChi2,LCm,ptail} and the implementation of machine learning techniques \cite{GNN,deepeaster,TomasPoS}. The SWGO collaboration is considering $\sim$1 $\text{km}^2$ layouts divided in multiple zones with different fill factors \cite{RubenLayouts,Petra2025ICRC,MPMT,FinalizingLayouts}: a core of radius 156 m with fill factor 60\%-70\% and one or more outer zones with fill factor $\leq5\%$. The collaboration aims to reach a 10\% energy resolution and $0.1^\circ$ angular resolution, improving performance and sensitivity with respect to the current generation of wide FoV observatories. This evaluation is still ongoing due to the limits in the statistics at high-energy \cite{swgo,UHEPerformance}.

To determine the array layout and the design of the detector (from its mechanics to the Cherenkov light sensors) for a wide field of view observatory (like SWGO will be), a full simulation (i.e. GEANT4-based code) of the experiment response for all candidate layouts is commonly required in conceptual design phases. The idea is to compare gamma/hadron separation, resolution and sensitivity of the layout candidates to determine the best option. Generally, experiments of this type use a two step approach. First, the CORSIKA \cite{corsika} software is used to produce a set of EAS at the observation altitude of the experiment. Only showers generated by gamma rays and protons are needed in a first approximation. Heavier atomic nuclei are not needed since they should be easier to recognise with respect to proton primaries. Then, a GEANT4-based \cite{geant4} simulation is used to simulate the detector response to the CORSIKA showers. GEANT4 tracks all particles from their starting position, adding all the Cherenkov and Bremsstrahlung photons produced during the propagation. Due to the huge amount of secondary particles in EAS (even tens of millions at a few PeV) and the photons produced in water, this second step requires huge computational times. It becomes difficult to obtain enough statistics to evaluate the g/h separation power, due to the amount of required proton events. Such an amount needs to be high enough to reliably test the strength of the used criteria. For instance, LHAASO \cite{LHAASOSeparation} reached a $10^{-4}-10^{-5}$ hadron rejection power.

To reduce the computation time of the full simulation codes, the authors have developed a parametrised simulation approach. A GEANT4-based code was used to study the tank response to single particles, obtaining a parametrisation of the tank response. Having such parametrisation, GEANT4 is no longer needed, and the response of the experiment to an EAS is then obtained by summing the response to each of its secondary particles. Since the parametrisation skips all the tracking operations performed by GEANT4, this results in a significant reduction of the computation time, which will be discussed in the last section of this article. Another group in the collaboration is applying our same concept to the simulation of a lake option for SWGO \cite{LakeSimSWGO}.

In this work, the authors describe the implementation of a fast simulation code for WCD arrays. However, the concept of this work is general and can be applied to other types of detectors and arrays. In principle, this approach should also work for experiments which detect the secondary particles with different particle detectors. First, we will describe what the look-up table (LUT) is, how it is possible to build it, and how to use it to provide a parametrisation of the tank response. Then, we will discuss part of the technical implementation of the code, the challenges we found, and the solutions we implemented to reach a good agreement with the full simulation. We will conclude with a description of the validation process we followed to prove this concept.

\section{Parametrisation of the tank response}
We define a hit as a particle touching the external walls of the detector. The parametrisation of a tank response (e.g., signal amplitude and time of the signal) is a model or distribution that describes the behavior of the tank as a function of the hit characteristics. All of this information must be stored inside a LUT that the simulation will use to replicate the tank response. In the next paragraph, the concept behind the preparation of a LUT will be described.

\subsection{Building a look-up table}
Considering a detector of any shape, the hits of the particles entering it can be described by seven variables:

\begin{itemize}
	\item Particle type
	\item Particle energy
	\item ($x$,$y$,$z$) of the hit: in the most generic case, the detector has no geometrical symmetry, and a complete map of the surface is needed to evaluate the detector response
	\item ($\theta$,$\phi$) angles of the particle direction: in the most generic case, all the possible angles should be considered, or at least the ones that lead the particle to enter the detector
\end{itemize}

Each possible hit is a point in this 7-dimensional parameter space. First of all, we can discretize it using a 7-dimensional grid and simulate (with a full GEANT4-based simulation) the detector response of the vertices. For each of those simulations, we can build histograms of the tank response variables, fit them, and write the parameters in a table with all the information. With this convention, each row of the table describes one of the simulations of the tank. Such table has $7+N_{res}$ columns:

\begin{itemize}
	\item 7 numbers to identify the coordinates of the simulated hit in the 7-parameter space.
	\item $N_{res}$ is the number of parameters needed by the fits of the tank response.
\end{itemize} 

This generic approach needs a very large number of simulations. Assuming 10 values for each variable, a complete LUT requires running $10^7$ simulations. But, in reality, any detector has some degree of symmetry that could be exploited. In the following paragraph, we describe, as an example, the procedure used to build the LUT of one of the WCD designs proposed for SWGO.

\subsection{Example: SWGO double-layer tank}
\begin{figure}
	\centering
	\begin{tikzpicture}
		\draw[thick,draw=black] (11,5.5) rectangle ++ (5.2,4.10) node [black,yshift=-4cm,xshift=-1.5cm] {};
		\draw[black, thick] (11,6.18) -- (16.2,6.18);
		\draw[black, line width=0.75mm] (10.5,5.5) -- (10.5,6.18) node[midway, left] {68 cm};
		\draw[black, line width=0.75mm] (10.5,6.18) -- (10.5,9.6) node[midway, left] {342 cm};
		\draw[black, line width=0.75mm] (10.4,5.5) -- (10.6,5.5);
		\draw[black, line width=0.75mm] (10.4,6.18) -- (10.6,6.18);
		\draw[black, line width=0.75mm] (10.4,9.6) -- (10.6,9.6);
		\draw[black, line width=0.75mm] (11,5) -- (16.2,5) node[midway,below] {520 cm};
		\draw[black, line width=0.75mm] (11,4.9) -- (11,5.1);
		\draw[black, line width=0.75mm] (16.2,4.9) -- (16.2,5.1);
		\draw[thick,fill=blue](13.6,9.6) circle (.2) node [black,yshift=.55 cm] {(0,0,H)};
		\draw[thick,fill=blue](16.2,9.6) circle (.2) node [black,xshift=1 cm] {(R,0,H)};
		\draw[thick,fill=blue](16.2,5.5) circle (.2) node [black,xshift=1 cm] {(R,0,0)};
		\draw[thick,fill=black](13.6,6.37) circle (.15) node [black,xshift=1 cm] {};
		\draw[thick,fill=black](13.6,6.14) circle (.1) node [black,xshift=1 cm] {};
	\end{tikzpicture}
    \caption{Design of the double-layer tank \cite{Ruben2LayerTank} used for developing and testing the fast simulation tool. The upper layer is equipped with an up-facing 10'' PMT while the lower one with a down-facing 8'' PMT.}
	\label{fig:tank-D}
\end{figure}
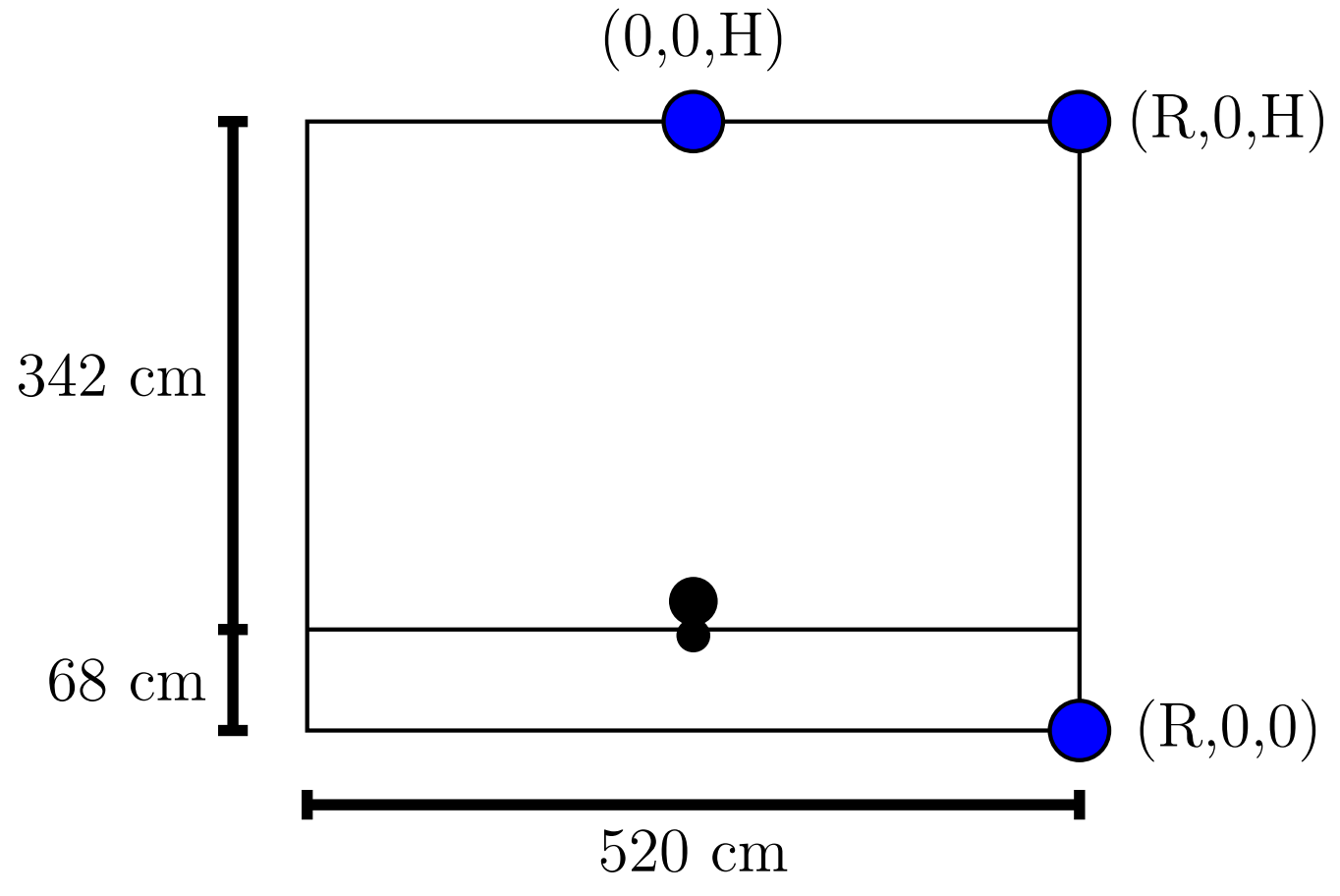

SWGO \cite{swgo2019wp,swgo} will measure the ground distribution of secondary EAS particles throu\-gh WCDs; one of the detector concept previously studied within SWGO-related design efforts is a double-layer tank \cite{Ruben2LayerTank,swgoDoubleLayerArticle}. The upper layer measures the electromagnetic component of the EAS and acts as a shield for the lower layer, making it more sensitive to the EAS muons. Considering a cylindrical tank as in Figure \ref{fig:tank-D}, with radius 260 cm and height 412.4 cm, we can significantly reduce the number of simulations by using the following variables:
\begin{enumerate}
	\item Particle type (PID)
	\item Particle energy (E)
	\item Zenith angle of the direction of the particle ($\theta$), measured starting from the downward z-axis (a down-going particle has $\theta=0^\circ$).
	\item Azimuth angle of the direction of the particle ($\phi$), measured starting from the axis defined by the center of the tank and the hit position (a particle moving towards the tank center has $\phi=180^\circ$).
	\item Distance of the particle hit from the central axis of the tank ($d_R$)
	\item Distance of the particle hit from the base of the tank ($d_H$)
\end{enumerate}

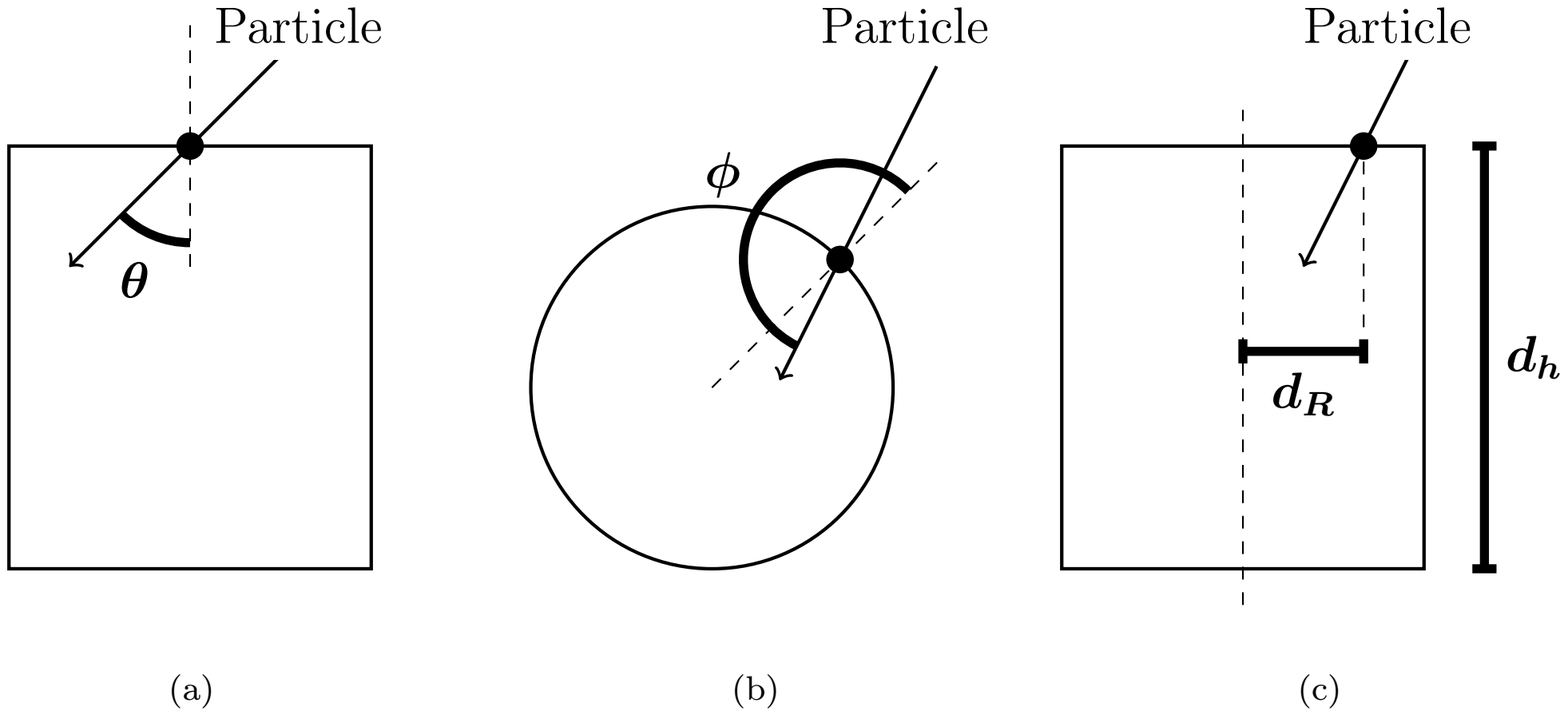
\begin{figure}
	\centering
	\begin{subfigure}{0.3\linewidth}
		\centering
		\begin{tikzpicture}
			\draw[thick,draw=black] (11,5.5) rectangle ++(3,3.5) node [black,yshift=-4cm,xshift=-1.5cm] {};
			\draw[black, thick,->] (13.5,10) -- (11.5,8);
			\node[text width=1.2cm,fill=white] at (13.3,10) {Particle};
			\draw[black,dashed] (12.5,8) -- (12.5,10);
			\coordinate (A) at (12.5,8);
			\coordinate (B) at (12.5,9);
			\coordinate (C) at (11.5,8);
			\pic [draw, line width=0.75mm,"$\bm{\theta}$", angle eccentricity=1.5,angle radius=.8cm] {angle =  C--B--A};
			\draw[thick,fill=black](12.5,9) circle (.1);
		\end{tikzpicture} 
		\caption{}
	\end{subfigure}
	\begin{subfigure}{0.3\linewidth}
		\centering
		\begin{tikzpicture}
			\draw[thick,fill=none](0,0) circle (1.5) node [black,yshift=-2cm] {};
			\draw[thick,fill=black]({sqrt(2)*0.75},{sqrt(2)*0.75}) circle (.1);
			\draw[black,dashed] (0,0) -- (1.9,1.9);
			\draw[black, thick,->] ({sqrt(2)*0.75+.8},{sqrt(2)*0.75+1.6}) -- ({sqrt(2)*0.75-.5},{sqrt(2)*0.75-1});
			\node[text width=1.2cm,fill=white] at (1.5,3) {Particle};
			\coordinate (A) at ({sqrt(2)*0.75-1},{sqrt(2)*0.75-2});
			\coordinate (B) at ({sqrt(2)*0.75},{sqrt(2)*0.75});
			\coordinate (C) at (2.5,2.5);
			\pic [draw,line width=0.75mm, "$\bm{\phi}$", angle eccentricity=1.5,angle radius=.8cm] {angle =  C--B--A};
		\end{tikzpicture}  
		\caption{}
	\end{subfigure}
	\begin{subfigure}{0.3\linewidth}
		\centering
		\begin{tikzpicture}
			\draw[thick,draw=black] (11,5.5) rectangle ++(3,3.5) node [black,yshift=-4cm,xshift=-1.5cm] {};
			\draw[black, thick,->] (14,10) -- (13,8);
			\node[text width=2cm,fill=white] at (14,10) {Particle};
			\draw[thick,fill=black](13.5,9) circle (.1);
			\draw[black, line width=0.75mm] (14.5,9) -- (14.5,5.5) node[midway, right] {$\bm{d_h}$};
			\draw[black, line width=0.75mm] (14.4,9) -- (14.6,9);
			\draw[black, line width=0.75mm] (14.4,5.5) -- (14.6,5.5);
			\draw[black,dashed] (12.5,5.2) -- (12.5,9.3);
			\draw[black, line width=0.75mm] (12.5,7.3) -- (13.5,7.3) node[midway, below] {$\bm{d_R}$};
			\draw[black, line width=0.75mm] (12.5,7.2) -- (12.5,7.4);
			\draw[black, line width=0.75mm] (13.5,7.2) -- (13.5,7.4);
			\draw[black,dashed] (13.5,7.5) -- (13.5,9);
		\end{tikzpicture} 
		\caption{}
	\end{subfigure}
	\caption{Definitions of the hit coordinates/angles. (a) Zenith angle $\theta$ (b), Azimuth angle $\phi$ (c), Hit radius $d_R$ (note that the distance is not the projection in the 2D plane but is calculated in the 3D space), and height $d_h$}  
	\label{fig:hit-definitions}
\end{figure}   

Figure \ref{fig:hit-definitions} shows a visual representation of the defined angles and coordinates. Using these variables, we can exploit the cylindrical symmetry of the tank. The number of spatial variables is reduced to two and to evaluate the dependence from the hit position is necessary only to simulate hits over two segments (as shown in Figure \ref{fig:tank-D}): from (0,0,H) to (R,0,H) and from (R,0,H) to (R,0,0). Where R and H are respectively the radius and the height of the tank. The number of simulations can be further reduced according to the following considerations based on the symmetry of the detector:
\begin{enumerate}
	\item The main contribution of the signal is due to $\gamma$, $e^-$, $e^+$, $\mu^-$, and $\mu^+$, but since the detector measures the Cherenkov emission of the particles, it is only needed to simulate $\gamma$, $e^-$, and $\mu^-$. Their corresponding anti-particles will have the same behavior.
	\item The energy range and the number of bins can be optimized for each particle type. From CORSIKA simulations, we obtained the energy distribution of the EAS particles in the showers generated by primaries in the energy range of interest (e.g., in the UHE range case 100 GeV$\leq$E$\leq$10 PeV). Electrons and photons have similar energy distributions (from 1 MeV to 10 GeV), while muons extend to greater energies (300 MeV to 100 GeV). Furthermore, the amount of Cherenkov radiation is mainly dependent on how much material is crossed by the particles, so we expect a greater energy dependence for electrons and photons. We decided to study 3 values per decade for photons/electrons and 2 for muons.
	\item For particles entering from the ceiling of the tank, only azimuthal angles from $0^\circ$ to $180^\circ$ should be considered, since the other values can be obtained from simulations for $\phi$ specular with respect to $180^\circ$ (e.g. $120^\circ$ and $240^\circ$).
	\item For side-entering particles, the same idea applies and only the range [$120^\circ$,$180^\circ$] needs to be simulated. Particles with a smaller angle would not enter the detector or would cross too little water to produce a signal.
	\item There is no need to vary $\phi$ for particles which enter from the center of the cover of the tank, due to the cylindrical symmetry.
\end{enumerate}

\subsubsection{Tank simulations}
We produced the simulations for the LUT of the double-layer tank using the full simulation software of the collaboration (HAWCSim \cite{hawc}), used here as a reference for validation purposes. Furthermore, we followed this scheme:

\begin{itemize}
	\item For electrons and photons, we simulated energies of: 1 MeV, 3 MeV, 7 MeV, 10 MeV, 30 MeV, 70 MeV, ..., 7 GeV, and 10 GeV
	\item For muons, we simulated energies of: 0.3 GeV, 1 GeV, 3 GeV, 10 GeV, 30 GeV,  100 GeV, and 300 GeV
	\item Zenith angle was varied from 0$^\circ$ to 80$^\circ$ at steps of 10$^\circ$
	\item For particles entering from the ceiling of the tank: \begin{itemize}
		\item Fixed hit $z$ at 412.4 cm
		\item 10 hit positions, along the tank radius from 0 cm to 260 cm
		\item Azimuth angle from 0$^\circ$ to 320$^\circ$ at steps of 40$^\circ$, plus and an additional bin for 180$^\circ$
	\end{itemize}
	\item For particles entering from the side of the tank: \begin{itemize}
		\item Fixed hit radius at 260 cm
		\item 8 hit z positions from 0 cm to 412.4 cm (extremes discarded)
		\item Azimuth angle varied from 120$^\circ$ to 240$^\circ$ at steps of 30$^\circ$
	\end{itemize}
\end{itemize}

From our tests, this set is enough to reproduce the tank response, even if a couple more energy bins for electrons and photons would have helped in reproducing the most energetic secondary particles in UHE showers. This is based on the results of the validation step that will be discussed in Section 4.

\subsubsection{Tank response}
The most difficult step in the LUT production is choosing the best distributions to fit the tank response. For each variable to reproduce (e.g., signal amplitude and arrival time), it is necessary to find a single function fitting the simulated data. Then, for each simulation, the histogram of the variable must be fitted. For our test tank, we had to reproduce two types of variables: the number of photo-electrons (PEs) hitting the two PMTs, and the arrival time of the first PE hitting the upper PMT. It is not necessary that the distribution parameters have a physical interpretation (even if it would be advisable) since the objective of the parameterization is to reproduce the tank response.

\paragraph{Number of PEs}
In the fitting procedure, we performed a minimum $\chi^2$ minimization. For the binning of the histograms, we had to implement a flexible criterion. We considered the maximum and minimum value of the data we had to fit, and we tried to divide that range into 100 bins. The distribution we aim to fit is defined only on integers, hence we adjusted the bin size to center all the bins at integer values. For the errors, we considered poissonian fluctuations of the number of entries of the histograms.

We found that for most simulations, we can describe the number of PEs arriving at a PMT with the negative binomial distribution from the \texttt{scipy} \cite{scipy} Python package:

\begin{equation}
	f(k;n,p)=\binom{k+n-1}{n-1}p^n(1-p)^k\ \ \ \ \text{for}\ k\geq0\ \text{and}\ p\in(0,1]
\end{equation}

This is the distribution of the number of failures (k) that are obtained before getting $n$ successes in repeated trials of a binomial process with probability $p$. The reason we choose this model is that it can be seen as a poisson-like distribution, with an additional parameter describing a longer tail. In most of the simulations, this is enough, but there are certain cases in which the distribution is more complex.

\begin{figure}
	\centering
	\includegraphics[width=0.95\linewidth]{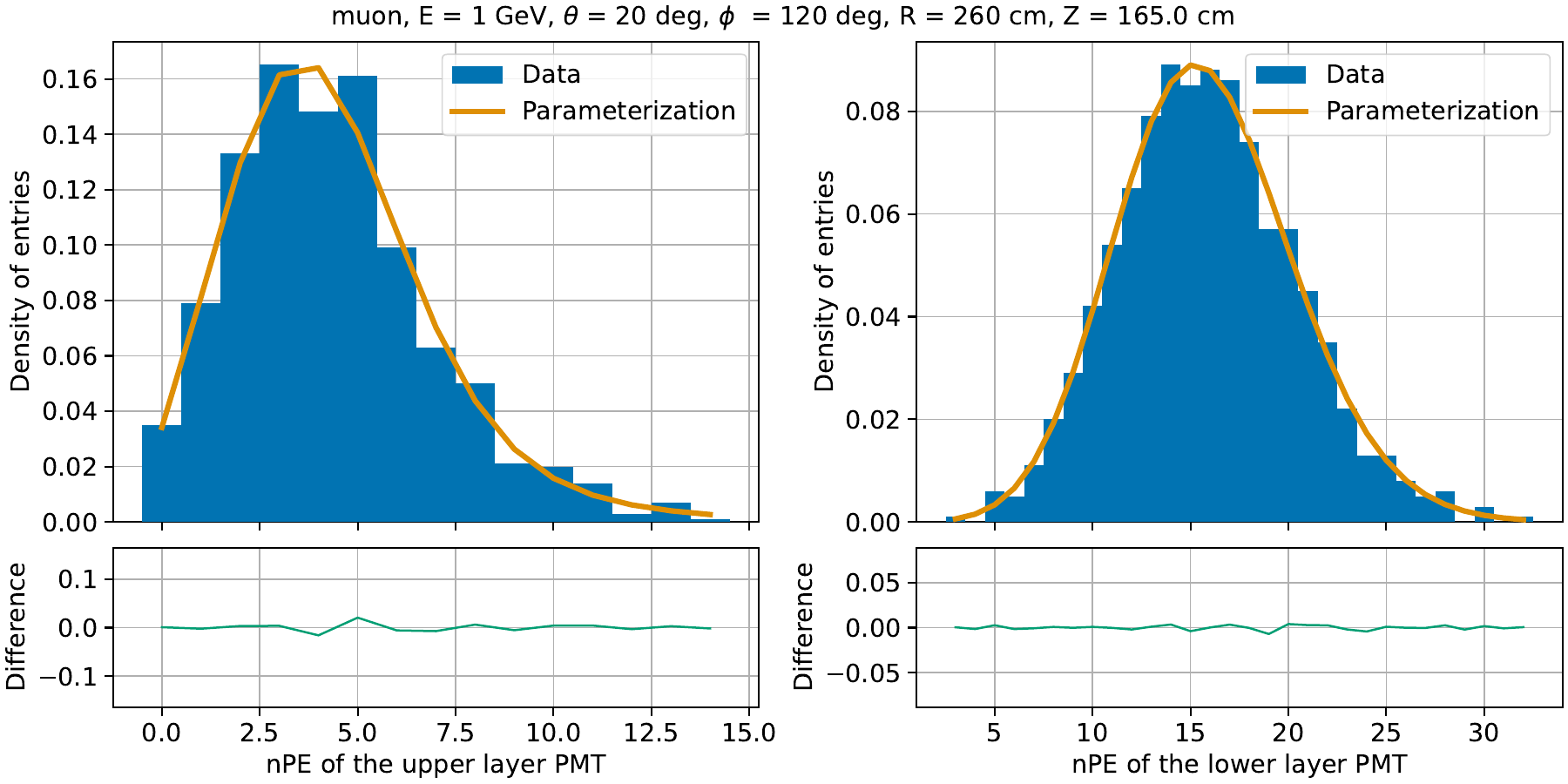}\\
	\includegraphics[width=0.95\linewidth]{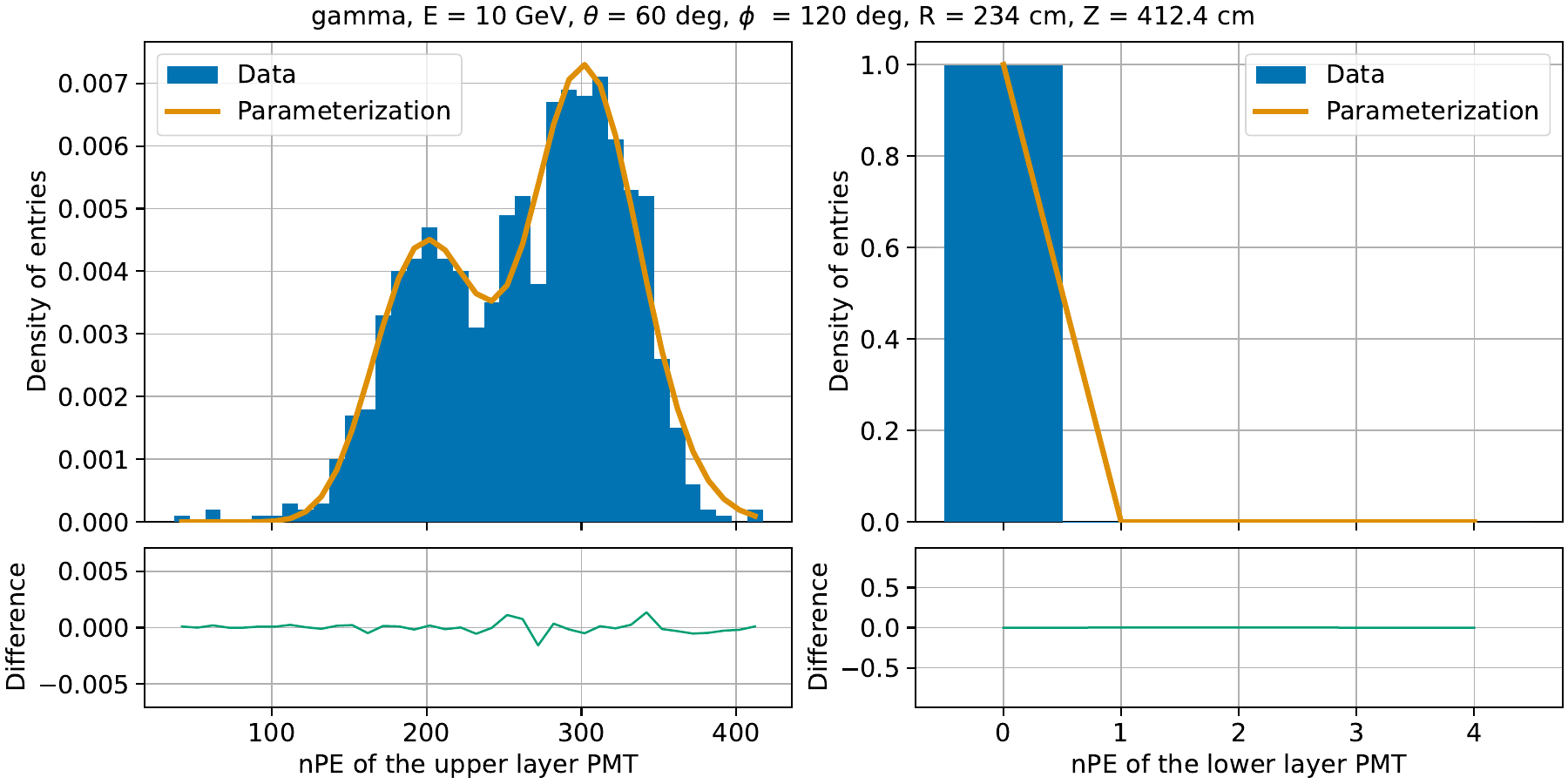}
	\caption{Signal (and relative parameterization) of the tank in response to different parti\-cles. Each row contains the parametrisation of the response of the 2 PMTs of the tank to the same particle.}
	\label{fig:nbinom}
\end{figure}

We noticed that the most irregular points of the LUT are low-energy particles or particles that cross a small amount of water. The difficulty in the fit can be due to two reasons:
\begin{itemize}
	\item The histogram is strongly peaked at 0 and only a few counts are in the right tail. In these cases, it is difficult to fit the right tail due to the uncertainty generated by the small amount of counts (as shown in the bottom right plot of Figure \ref{fig:nbinom}). Furthermore, it is possible that the signal is always 0 for geometrical reasons (e.g., a particle not crossing the lower layer will not leave a signal in the lower PMT).
	\item The histogram contains multiple peaks, as shown in the lower left plot of Figure \ref{fig:nbinom}. In this case, a single negative binomial will never be able to describe the data correctly.
\end{itemize}

As a solution, we decided to model the signal as the linear combination of 2 negative binomials using:

\begin{equation}
	g(k) = \alpha\cdot f(k,n,p) + (1-\alpha)\cdot f(k;n',p')
	\label{eq:linear-combination}
\end{equation}

Where $\alpha$ is limited to the interval [0,1] to preserve the normalization of the distribution. Figure \ref{fig:nbinom} shows some example simulations superposed with the fitted parameterizations.

\paragraph{Tank time}
For this variable, we followed a more qualitative approach. We always divided the interval covered by the data into 100 bins, cutting the distribution at 100 ns. We did not consider errors in the fitting procedure, but at any point in our analysis, the checks we performed suggested the need for a more detailed check. Although for future LUTs we plan to follow the same procedure we took for the tank signal.

\begin{figure}
	\centering
	\includegraphics[width=0.49\linewidth]{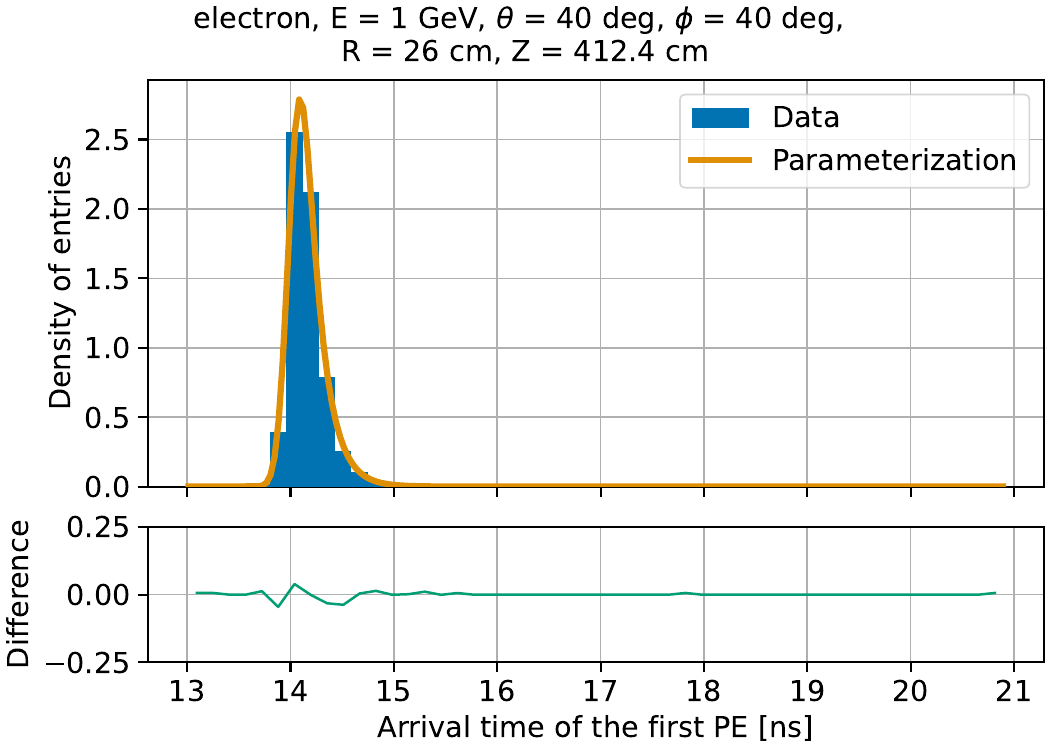}
	\includegraphics[width=0.49\linewidth]{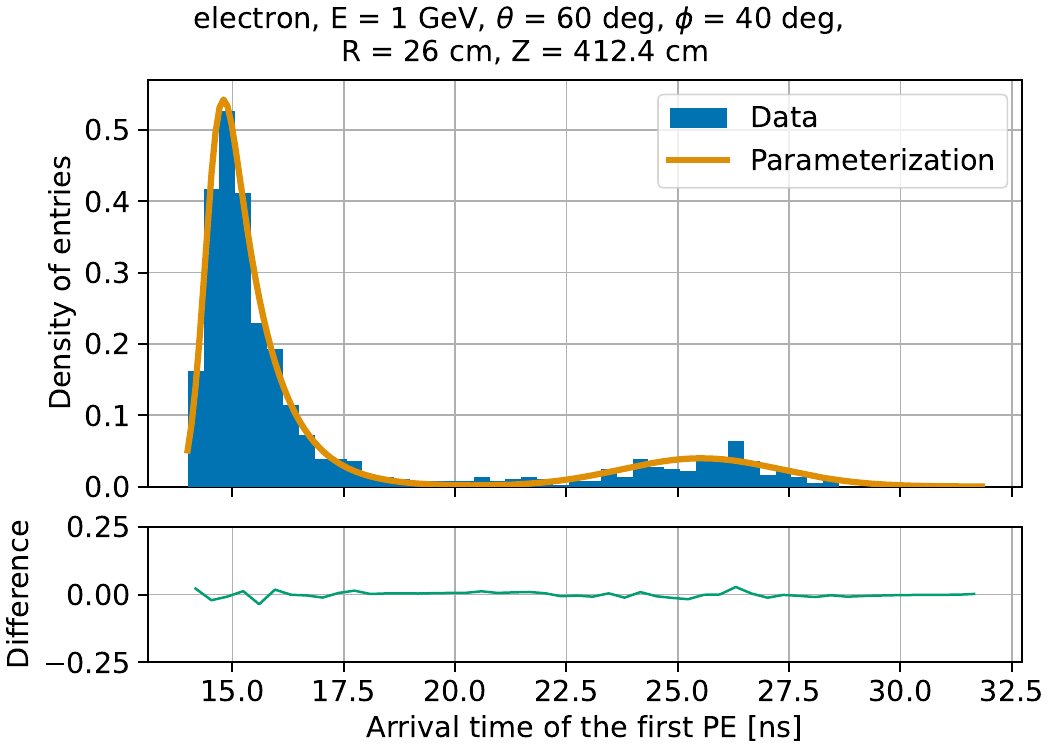}
	\caption{Time response (and parameterization) of the tank to different particles.}
	\label{fig:double-exponnorm}
\end{figure}

For the time variable, often the histograms have multiple peaks as shown in the right plot of Figure \ref{fig:double-exponnorm}. Most of the time, one peak is higher than the others, but sometimes the second one can not be neglected. The single peaks can be fitted with an exponentially modified gaussian from the \texttt{scipy} package:

\begin{equation}
	\text{expnorm}(x;\mu,\sigma,\lambda)=\frac{\lambda}{2}\text{exp}\Big(\frac{\lambda}{2}(2\mu+\lambda\sigma^2-2x)\Big)\text{erfc}\Big(\frac{\mu+\lambda\sigma^2-x}{\sqrt{2}\sigma}\Big)
	\label{eq:exponnorm}
\end{equation}

\noindent where erfc(x) is defined as

\begin{equation}
	\text{erfc}(x)=\frac{2}{\sqrt{\pi}}\int_x^\infty e^{-t^2}dt
\end{equation}

\noindent To describe the multiple peaks, we defined the parametrisation of the time variable as a linear combination of two such distributions, similarly to what we did for the nPE distribution in equation \ref{eq:linear-combination}.

The reason behind the presence of multiple peaks is due to the reflective walls of the tank. PEs that arrive directly at the PMT will travel less time in water with respect to the ones that arrive after 1 or more reflections. A more detailed discussion on this is out of the scope of this article since the objective of the fitting step is to reproduce the tank response. Figure \ref{fig:double-exponnorm} shows some example distributions of the time variable and their relative parametrisation.

Lastly, we also added a minimum time parameter, equal to the smallest time of the simulation. We noticed that sometimes this model failed to reproduce the strong cut in the left tail of the distribution with the earliest peak. Basically, when using the LUT to reproduce the tank response, if the extracted time is smaller than the minimum one, another value is extracted until a valid one is obtained.

\subsubsection{Fit quality of the nPE distributions}
Since g/h separation relies much in the fluctuations of the signal amplitude, we followed a more strict procedure for the nPE distribution fits. Although a $\chi^2$ variable can not be properly computed for all the fits:
\begin{enumerate}
	\item There are 5 parameters in the fitted distribution, but certain histograms have a similar or even smaller number of bins. In these cases, we cannot perform a fit of the distribution, but the operation reduces to an interpolation. This happens only for low-energy electrons/photons or for particles that cross only a small part of the water (e.g., a few cm)
	\item A proper $\chi^2$ test would require at least 5 entries in each fitted bin, which does not always happen
\end{enumerate}

\begin{figure}
	\centering
	\includegraphics[width=0.48\linewidth]{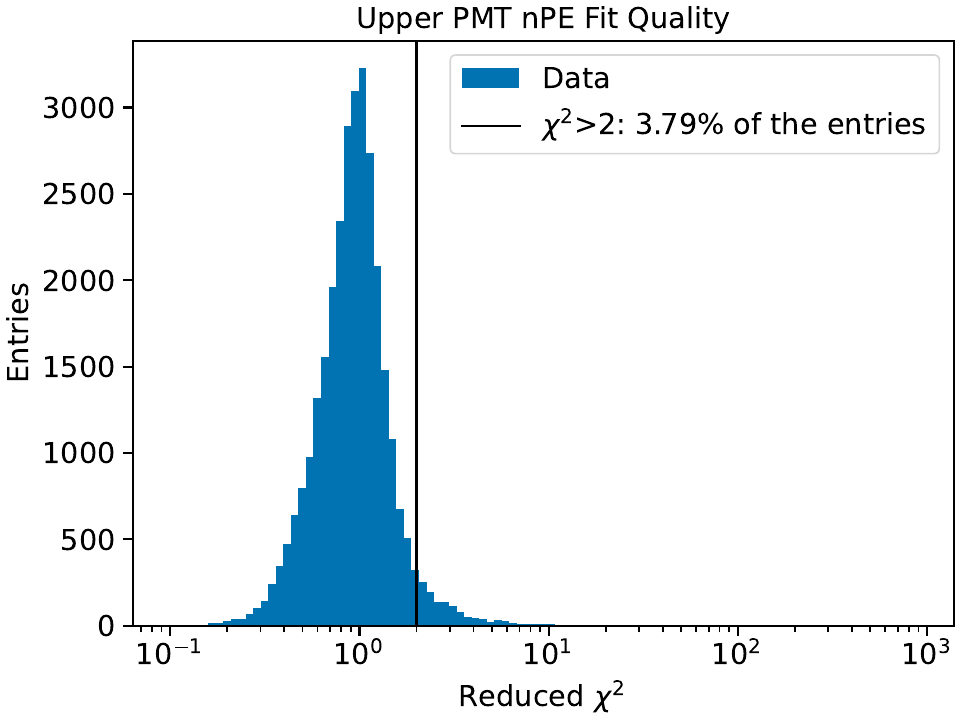}
	\includegraphics[width=0.48\linewidth]{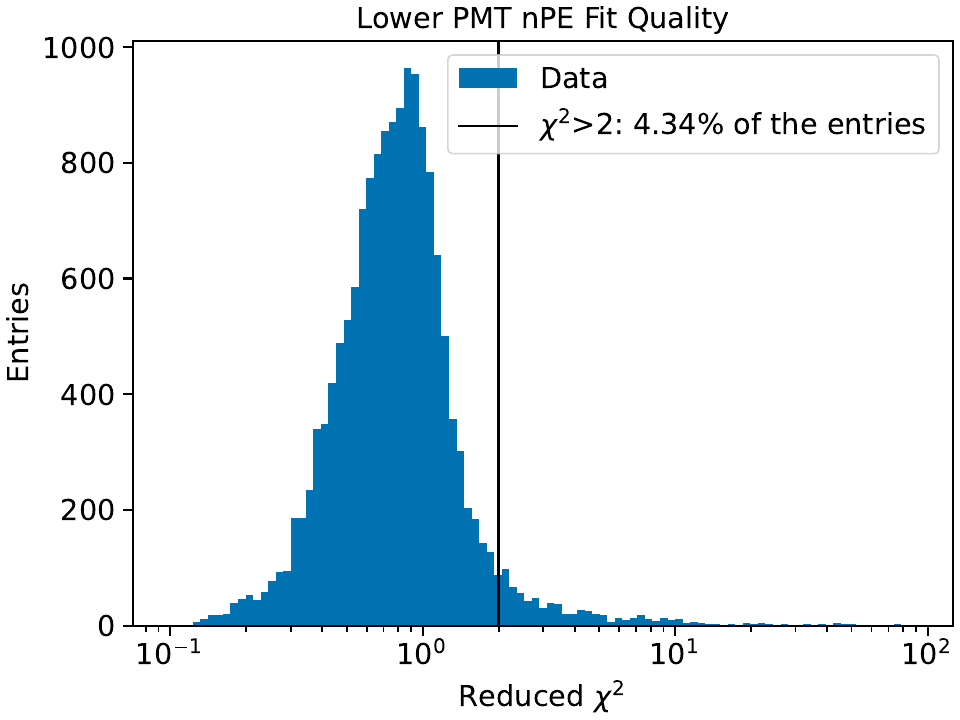}
	\caption{Histogram of the reduced $\chi^2$ of the fits of the tank signal for the upper PMT (left) and the lower PMT (right)}
	\label{fig:fit-quality}
\end{figure}

Nevertheless, it is possible to use the reduced $\chi^2$ variable to check if the model is correctly reproducing the data. In this sense, we calculated such a variable for all histograms with at least two degrees of freedom; the results of such analysis are shown in Figure \ref{fig:fit-quality}. For both PMTs, the average of the reduced $\chi^2$ distribution is around 1, and only $<$5\% of the fits are above 2 reduced $\chi^2$. We conclude that the chosen model is able to describe the tank response.

Another small challenge imposed by the procedure is the overparametrisation of the fit. In many cases, only a single peak can describe the whole distribution; this leads the minimizer to utilize the second one to describe fluctuations of the histogram generated by the limited amount of counts. This can generate quite some bad effects on the LUT if this happens in the long tails of these histograms. To prevent this, we found that disabling peaks describing less than 5\% of the counts was sufficient. This, reduces the goodness of some of the fits but removes such artifacts introduced by the minimizer.

\subsubsection{Look-up table regularization}
The simulation code will not be able to recognize periodic variables or detector specific symmetries (e.g., cylindrical symmetry). All the simulations skipped due to the exploitation of the tank symmetry must be described in the LUT file. We chose this approach to not hard-code these properties in the interpolator of the simulation and make the code more flexible to different detectors (with less symmetry). For example, this means that all information of $\phi=0^\circ$ was used to describe $\phi=360^\circ$. This last step of the LUT preparation is very important, because it is the step that lets us reduce the number of required simulations (see the discussion above).

\subsubsection{A couple of remarks about the LUT construction}
We used this example double-layer tank to illustrate the idea of the LUT and the procedure we used to build it. The LUT is essential to run this algorithm, but it is not part of it; it is a configuration file. Hence, all the models we chose and the fitting procedure we found are arbitrary and can be changed without rewriting the whole simulation from scratch. This is an important feature since describing different detectors could lead to big changes in the LUT. For example, adding more PMTs or tank response variables.

We now move to the description of how the simulation is performed and how all the information of the LUT is used to reproduce the tank response. We will return to this specific LUT in the validation section, in which we will compare the simulation of CORSIKA events using this LUT with the results obtained with the full simulation software (GEANT4-based) of the SWGO collaboration.

\section{Implementation of the simulation}
The fast simulation code can be described by the following procedure:
\begin{itemize}
	\item Setup: the CORSIKA file is opened, the configuration file is parsed, the random extraction object is instanced, the array, and the LUT are loaded and the interpolators are created.
	\item Loop on all events:
	\begin{enumerate}
		\item The core position of the event is extracted
		\item All particles of the event are initialized (e.g., shifted by the core position, rotated into the experiment reference system)
		\item Particles that are too distant from the array are discarded, also the ones with energy lower than the Cherenkov effect threshold (in water) are removed
		\item All hits with tanks of the array are evaluated
		\item The interpolator is used, and to extract the detector response
		\item The event output is written
	\end{enumerate}
	\item The output file is written, and the simulation ends
\end{itemize}

\subsection{Evaluating intersections}
We approximated the propagation of the particles with straight lines. Therefore, the evaluation of the intersections can be performed by solving the system of equations of the intersection of a line and a cylinder in a 3-dimensional space: 

\begin{equation}
	\begin{cases}
		x = p_xs+x_0\\
		y = p_ys+y_0\\
		z = p_zs+z_0\\
		z \geq z_{tank}\\
		x^2+y^2=r_{tank}^2\ if\ z < z_{tank} + h_{tank}\\
		x^2+y^2\leq r_{tank}^2\ if\ z = z_{tank} + h_{tank}
	\end{cases}
    \label{eq:intersection-system}
\end{equation}
Where $p_i$ are the 3 components of the particle momentum, $x_0$, $y_0$ and $z_0$ are the starting coordinates of the particle, $z_{tank}$ is the z coordinate of the base of the tank and $r_{tank}$ and $h_{tank}$ are respectively the radius and the height of the tank. The $s$ parameter is a dummy variable that is used to parametrize a line in a 3D space, it has the dimension of a distance over a momentum, and with equation \ref{eq:s-to-time-conversion} can be transformed into the time that the particles need to reach the hit position.

\begin{equation}
	t = \frac{s\cdot p}{\beta\cdot c}
	\label{eq:s-to-time-conversion}
\end{equation}

In equation \ref{eq:s-to-time-conversion}, $p$ is the momentum of the particle and $\beta$ the relativistic velocity. The system in equation \ref{eq:intersection-system} can also be used to perform a first reduction of the number of particles to track. It is necessary to define a cylinder that contains the whole detector array. Any particle that does not cross this cylinder will be unable to enter one of the tanks, and therefore it can be discarded. This is the most time intensive step of the simulation, since it is required to loop over all detectors to evaluate the intersections. 

For experiments that do not use cylindrical detectors, the equations used need to be adapted to the geometry of the tank (e.g., a cubic detector). Another option, which was not tested, could be to inscribe the non-cylindrical detector inside a cylinder and use that cylinder to evaluate intersections.

\subsection{Interpolator and output extraction}
The interpolator is the core object of the simulation; it uses an algorithm to find the correct tank response parameters. In our tests, we chose a nearest algorithm due to strong correlations in the fit parameters of the tank response. For our test tank, the input variables were PID, $E$, $\theta$, $\phi$, $d_R$, and $d_H$. For the interpolator, we used the one provided by \texttt{scipy}, with the \texttt{rescale} option set to true. This option morphs all the input variables to the [-1,1] interval in order to compensate for variables that change over several degrees of magnitude (e.g., particle energy). PID was not inserted in the interpolator, but we created a separate instance of the interpolator for each particle type. This was done to prevent a possible mix of information in the tank response.

We explored a linear interpolator, but this approach had three issues:
\begin{enumerate}
	\item The algorithm failed to describe the nonlinear correlations of the parameters of the tank response
	\item We noticed a great increase in the initialization time of the interpolation object (on our computer, it went from a few minutes to at least half an hour).
	\item The linear interpolator will return \texttt{NaN} values if the variables given as input are not in the volume of data of the interpolator (intuitively, if they are not surrounded by points of the LUT in the input parameter space). This is a feature of the \texttt{scipy} implementation, but using an extrapolation algorithm could lead to unpredictable results.
\end{enumerate}
About the third point, the nearest interpolator would instead return the nearest point it finds, without generating issues with \texttt{NaN} values.

When using a linear interpolator, we suggest a mixed approach: use the linear interpolator for all particle hits, and then substitute it with the output of a nearest interpolator if the first one returned a series of \texttt{NaN}. We found secondary particles up to several TeV in our test CORSIKA events. Although creating LUT points for those particles can be a challenge in storage and computing time, it is important to at least treat them in some way and not discard them simply because they have an energy greater than the highest simulated one.

\subsection{Particles that penetrate more than one tank}
\begin{figure}
	\centering
	\includegraphics[width=0.95\linewidth]{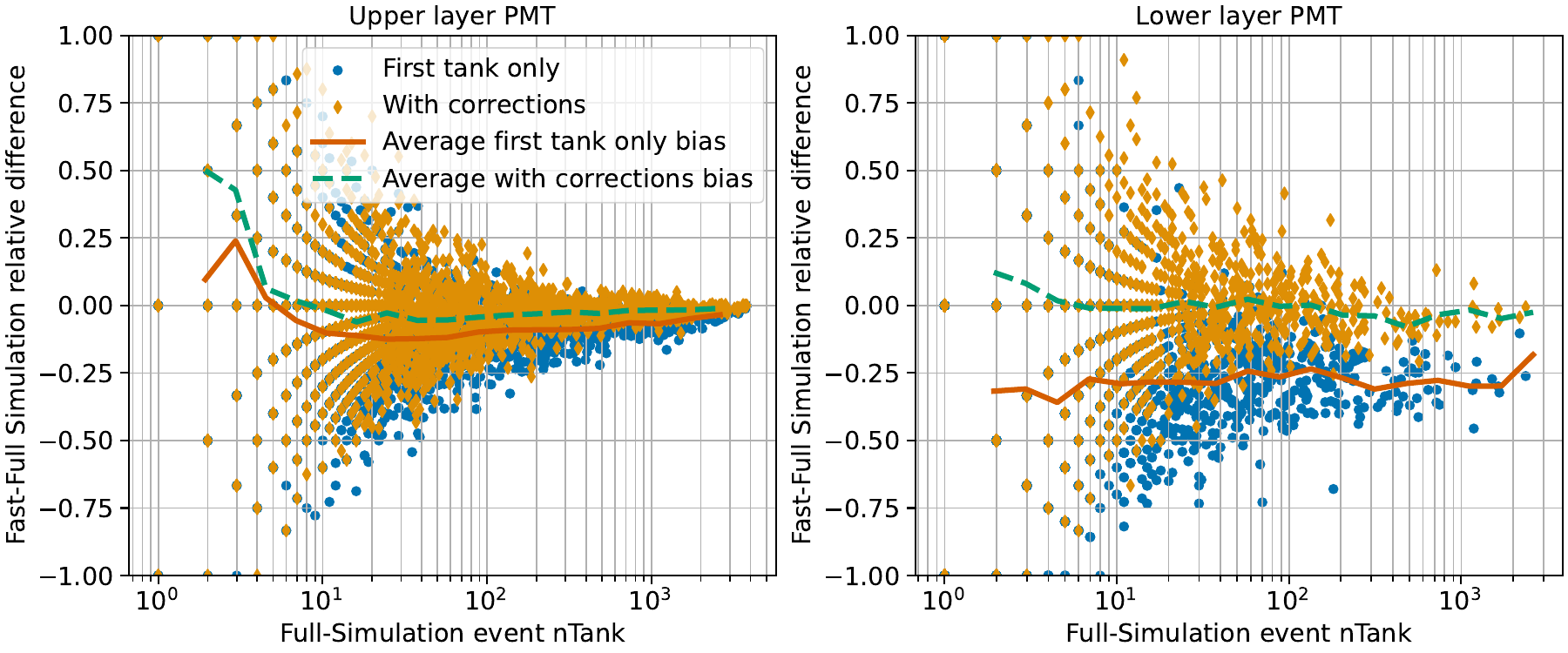}
	\caption{Relative difference of the number of tanks with at least 1 PE before and after adding the possibility for a particle to cross multiple detectors (comparing the same events simulated with both fast and full simulation). Left: only the upper layer was considered for hit detection; Right: only the lower chamber was considered for hit detection.}
	\label{fig:nhit-comparison}
\end{figure}

If the fill factor of the experiment is high, there is a high probability that the particles will cross multiple detectors. Initially, we tried to consider only the first hit (using time to discard the following ones). We soon realized that the result was not realistic. Figure \ref{fig:nhit-comparison} shows the number of tanks with at least 1 PE detected, which was much lower than the one predicted by a full GEANT4 simulation. The solution the authors found is particle-type dependent:
\begin{itemize}
	\item Muons (and anti-muons) are the easiest to handle, the tank response is stable enough with energy, and the losses in water are so small that all the intersections can be used without additional changes.
	\item Most of the electrons (and positrons) will lose energy mainly due to Brems\-strahlung. We neglected any direction change and used the formula:
	\begin{equation}
		E(L) = E_0e^{-L/X_0}
	\end{equation}
	where E(L) is the energy of the particle when leaving the tank, L is the amount of water crossed, and $X_0$ is the radiation length of water. The implementation of the average energy loss allows us use all the intersections associated with an electron with energy above the Cherenkov threshold.
	\item Photons are the most complicated particle to propagate in more than one tank. First of all, we refitted the whole photon section of the LUT after removing the contribution of the photons which entered the tank but did not interact inside it (which increases the number of events in which the signal is 0). Then, in the approximation that if the photon makes pair-production in the tank, none of the products will exit from it, we used the formula
	\begin{equation}
		P_{pp}(L)=e^{-\frac{7L}{9X_0}}
	\end{equation}
	to obtain the probability that the photon, entering the detector, interacts by pair production. At this point, it is possible to decide if a photon interacts in a detector by extracting a random number between 0 and 1. If the photon does not interact, we can propagate it to the next detector. Otherwise, we can extract the tank response and discard the following hits.
\end{itemize}

All of these solutions are approximations; they improve the result of the simulation, but only the full tracking of the particle would be able to reproduce the GEANT4-based simulation. Figure \ref{fig:nhit-comparison} shows the improvement in the number of tanks with at least 1 PE before and after these corrections, comparing both fast and full simulations. 

As expected, the lower layer of the tank is more influenced by this correction, as can be seen from the right panel of Figure \ref{fig:nhit-comparison} showing an underestimation of the number of hits of the bottom layer if only the first hit is considered. Before the corrections, some simulated events had half the amount of expected hits in the lower layer of the tanks, while with the corrections, the difference with the full simulation is within 10\%.

The most significant gain after applying this correction was observed in the muon component of the shower. Muons frequently cross multiple detectors, due to their low-energy loss, especially in the central region of the array, where the detectors are less spaced. This is especially important for the bottom layer of the tank.

\section{Validation}
\begin{figure}
	\centering
	\includegraphics[width=0.6\linewidth]{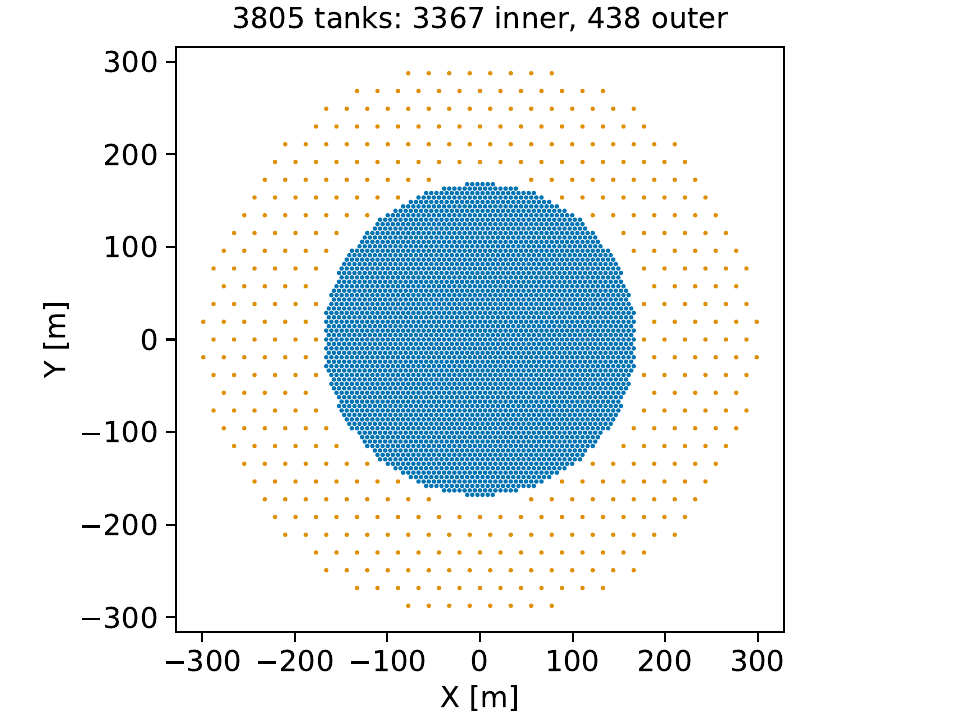}
	\caption{Array layout used for the validation of the fast simulation. It is divided into two zones: an inner core of 166 m radius with an 80\% fill factor, and an outer region of 300 m radius with a 5\% fill factor.}
	\label{fig:D1}
\end{figure}

In this section, the authors show some example events simulated with the fast simulation and compared with the full simulation framework used by the collaboration, serving as a reference. For this comparison, we used 2000 high-energy proton events (from 30 TeV to 1 PeV) using the test array shown in Figure \ref{fig:D1}. This array is divided into 2 zones: a central core of 166 m radius with 80\% fill factor and a lower density ring (5\% fill factor) of 300 m external radius. The core position of all events was extracted uniformly inside a 1500 m radius circle. We manually modified the distribution of the fast simulation cores in order to exactly reproduce the ones of the full simulation set, so that effects on the array structure do not influence our comparison.

For some variables, we will compare the prediction of both simulations using a relative ratio, using the following expression:
\begin{equation}
	\frac{X_{fast}-X_{full}}{X_{full}}
\end{equation}
where X is the variable we want to compare. For variables that are defined positive (e.g. total number of PEs), the relative ration cannot go below -1, which happens when the full simulation predicts a positive value while the fast simulation obtains 0.

\subsection{Event level variables}
\begin{figure}
	\centering
	\includegraphics[width=0.6\linewidth]{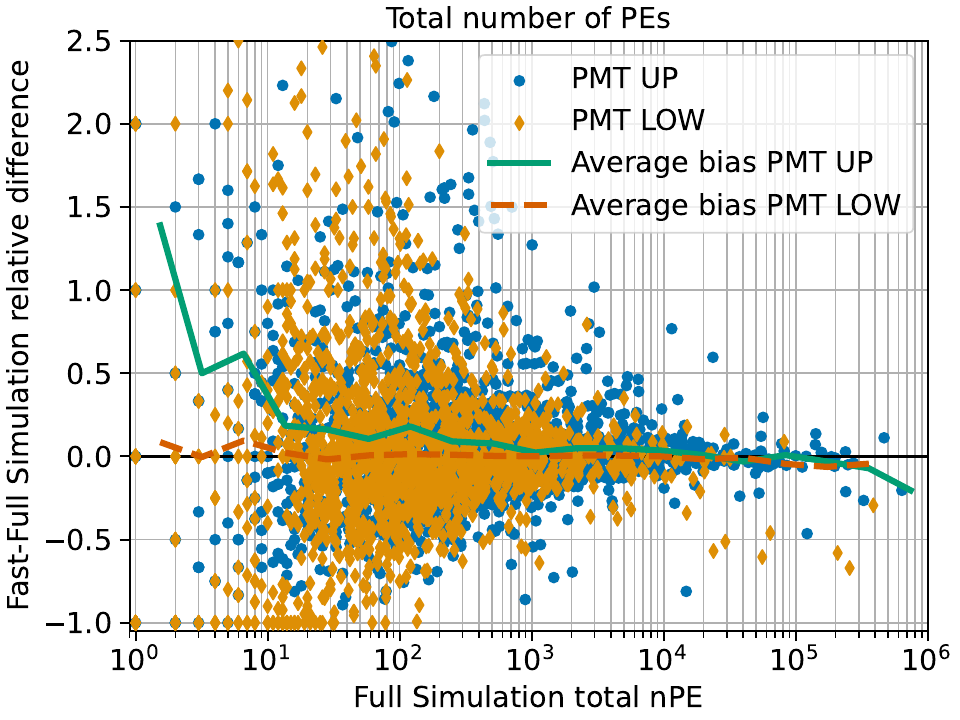}
	\caption{Difference of the total number of PEs detected by the array for 100 proton events thrown uniformly within a 1500 m radius from the center of the array. Different colors indicate which PMTs were used to calculate the number of PEs.}
	\label{fig:compare-variables}
\end{figure}

The first variables that we compared are the total number of PEs of the event and the number of tanks that detected at least one PE (see Figure \ref{fig:nhit-comparison}). Figure \ref{fig:compare-variables} shows the first one with different colors for the signal of the lower chamber and of the upper chamber. On average, the fast simulation slightly overestimates the number of total PEs; the discrepancy is not relevant in both layers staying under 100 PEs for events with a total signal under 2000 PEs. As the total signal of the event increases, the contribution of such difference to the total becomes more negligible.

We found events with a signal of the order of $10^4-10^7$ PEs in which the fast simulation underestimated the total signal by 50\%. We realized that the discrepancy was due to a small set of tanks very near the core. Those tanks were hit by particles with energy well above the maximum value we simulated for the LUTs. The easy way to fix this discrepancy is to add more energy bins, but we decided not to proceed due to 2 main reasons:
\begin{enumerate}
	\item As the energy of the particle grows, the size of the file produced to fit the LUT and the time needed to generate it grow.
	\item Electrons of 10 GeV are already able to saturate a PMT on their own and without the contribution of other particles. It is pointless to lose time optimizing the reproduction of data that a realistic reconstruction would not be able to process.
\end{enumerate}

\begin{figure}
	\centering
	\includegraphics[width=0.95\linewidth]{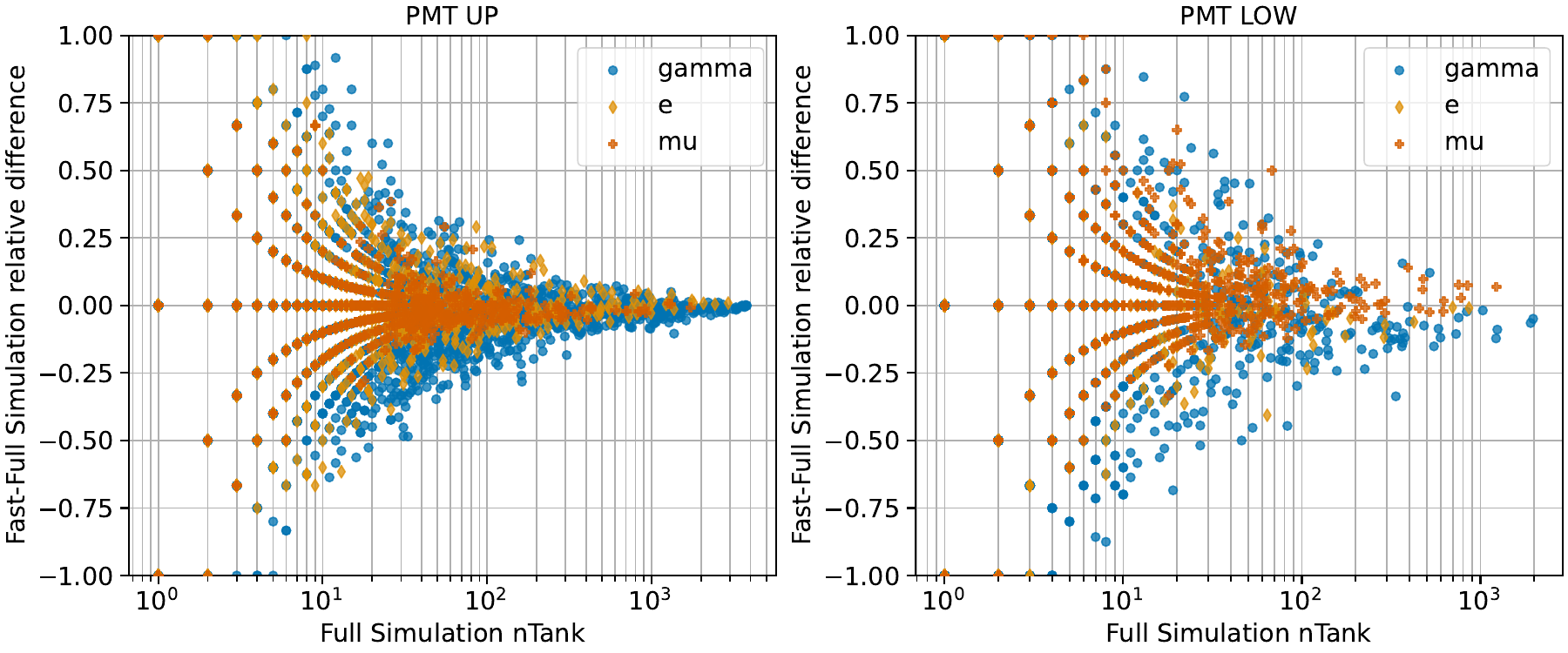}
	\caption{Comparison of the number of tanks triggered by particles (at least 1 measured PE). Left: only the upper tank layer is considered. Right: only the lower layer. The colors indicate the secondary particle types used in the tank hit computation.}
	\label{fig:secondary-hits}
\end{figure}

The number of tanks that detect a signal is slightly underestimated by the fast simulation in both layers. It is due to the approximations of the multiple hits. This can be checked by counting the number of hit tanks by a specific secondary particle. Figure \ref{fig:secondary-hits} shows the contribution of muons, electrons, and photons for both layers. As expected, the muons and electrons are well reproduced, while photons are reproduced with lower precision.

\subsection{Distributions of detected PEs in the shower plane}
\begin{figure}
	\centering
	\includegraphics[width=0.49\linewidth]{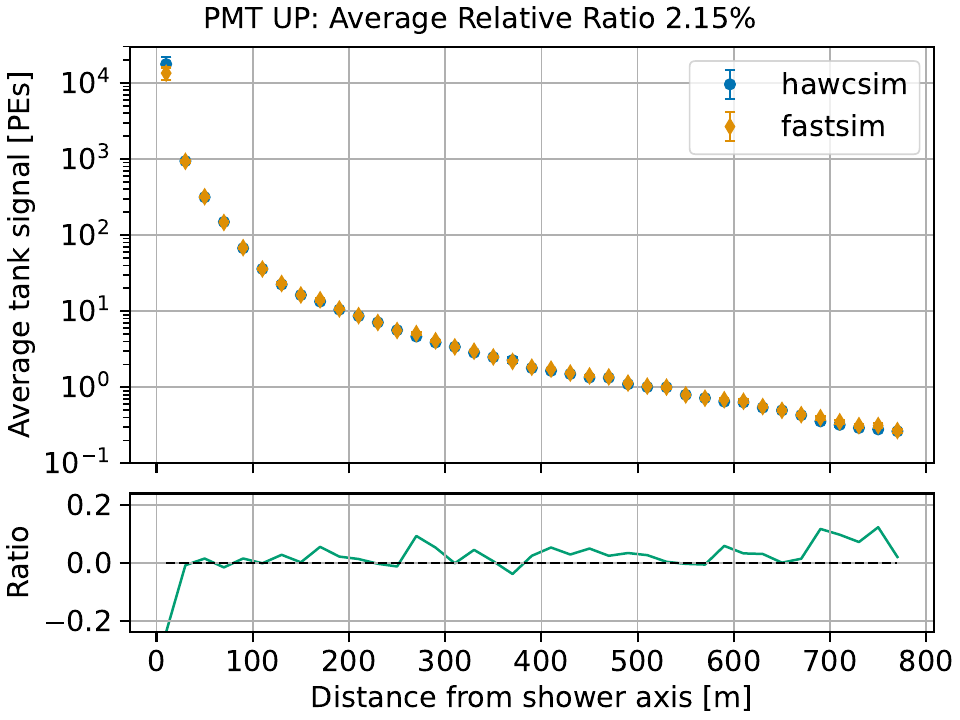}
	\includegraphics[width=0.49\linewidth]{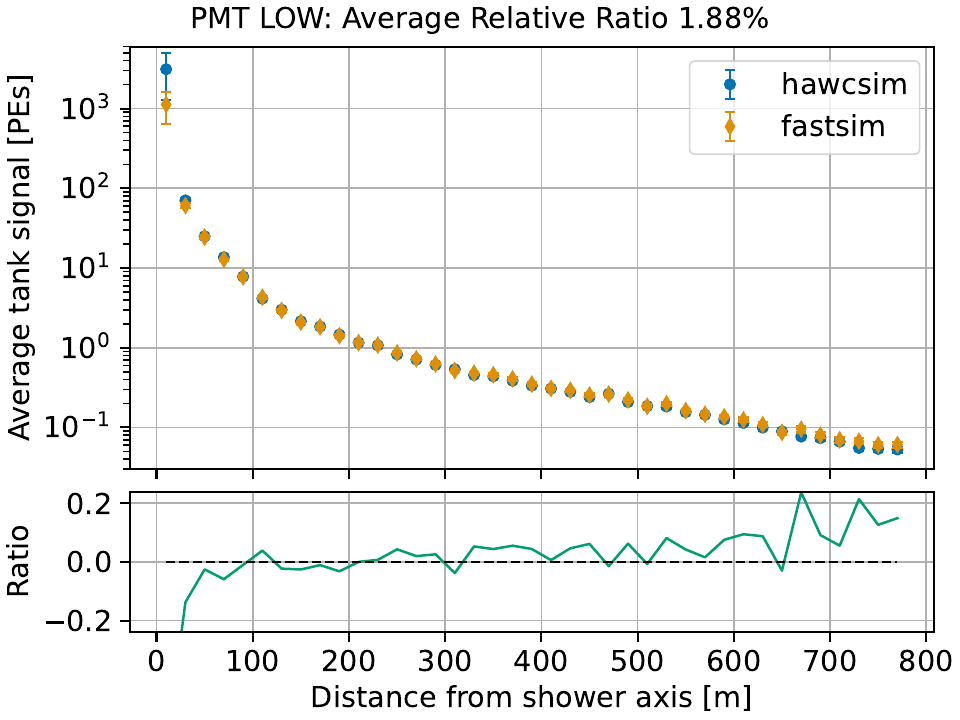}
	\caption{Average PE measured by the detector as a function of distance from the core for the upper-layer PMT (left) and lower-layer PMT (right).}
	\label{fig:lateral-distributions}
\end{figure}

By processing the signal of all PMTs, it is possible to produce a relation similar to the lateral distribution (density of particles of the EAS as a function of the distance from the shower axis). For each distance bin from the shower core, we computed the mean and standard deviation of the number of PEs. The results of such analysis are shown in Figure \ref{fig:lateral-distributions}.

As expected, the density of PEs is underestimated at small distances from the shower axis, due to the upper bound of the energy values in the LUT. Overall, the agreement between the number of PEs calculated by the full and the fast simulation is good for the tested energy range and configurations, showing that the latter correctly reproduces the detector behavior within the tested configuration. The relative difference of the lateral distribution oscillates around 0 for both layers and on average is $<$3\%. Although such a quantity is more unstable for the lower layer plot due to the smaller statistic, where the number of particles producing a signal is smaller. This is due to the smaller geometric acceptance and to the absorption of the upper chamber of the detector.

\begin{figure}
	\centering
	\includegraphics[width=0.49\linewidth]{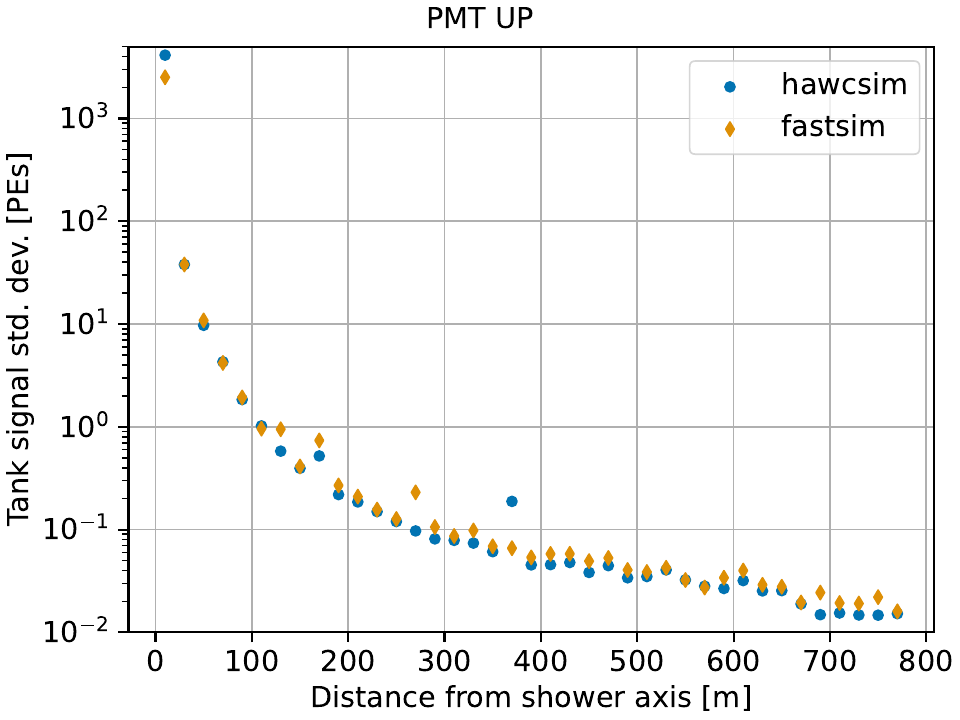}
	\includegraphics[width=0.49\linewidth]{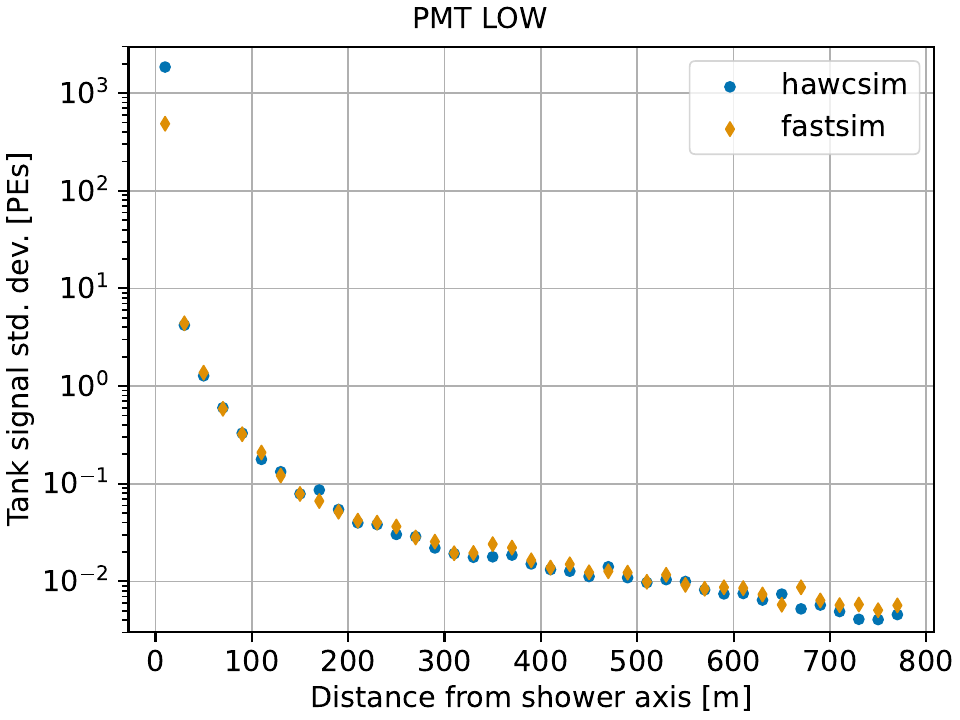}
	\caption{Standard deviation of tank signal as a function of distance from the core for the upper-layer PMT (left) and lower-layer PMT (right).}
	\label{fig:lateral-distributions-errors}
\end{figure}

We also inspected the reproduction of the fluctuations of the lateral distribution. Figure \ref{fig:lateral-distributions-errors} compares them in the two simulations. Also in this case, the quantity is consistent within the fast and full simulation. The only exception we found is in the first 20 m, where the fast simulation seems to underestimate the fluctuations. If one performs the same calculations only using the tanks in the sparse area of the array, the discrepancy vanishes. This indicates that the difference is due to the approximations done to simulate particles crossing multiple tanks.

\begin{figure}
	\centering
	\includegraphics[width=0.48\linewidth]{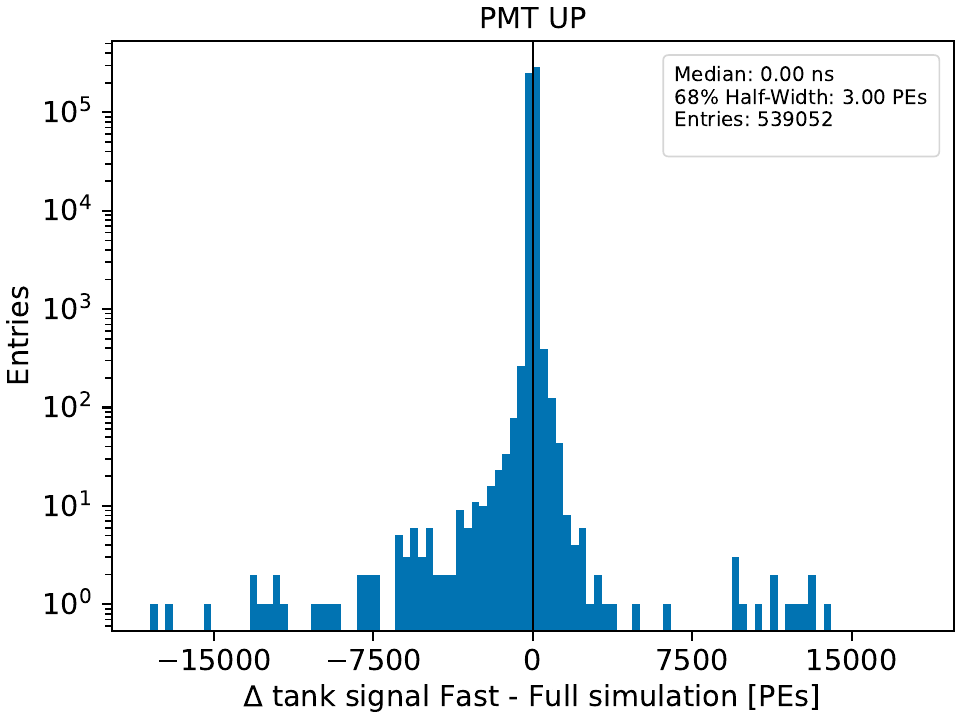}
	\includegraphics[width=0.48\linewidth]{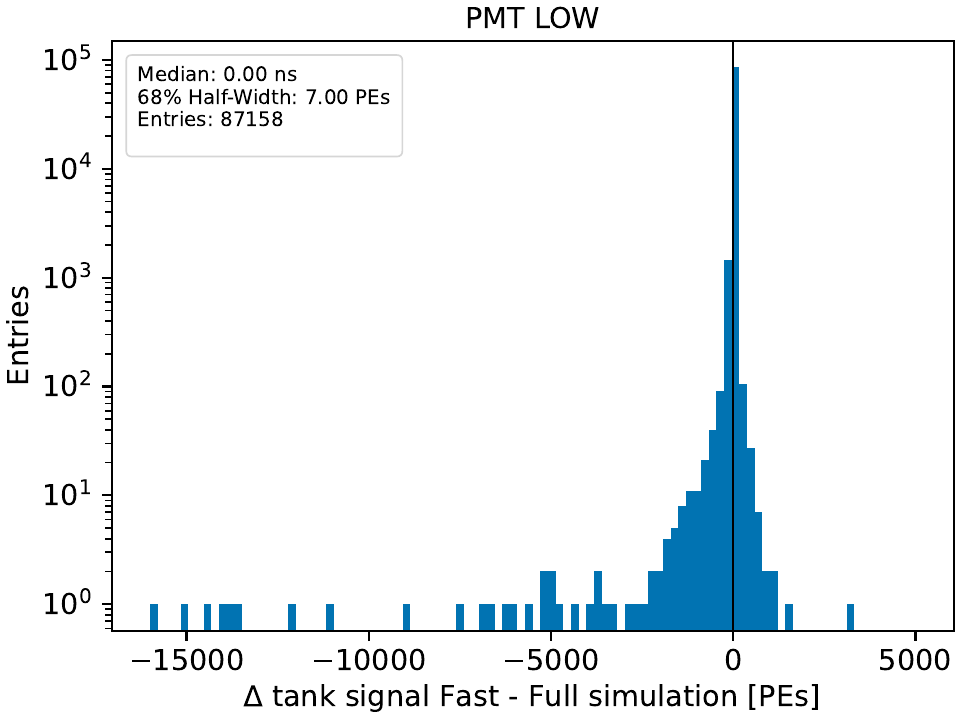}
	\caption{Difference of the signal in all the tanks which detected at least 1 PE in either fast or full simulation for upper-layer PMT (left) and lower-layer PMT (right).}
	\label{fig:delta-tank-signal}
\end{figure}

It is also possible to check the difference of the tank signal predicted by both fast and full simulation algorithms. Figure \ref{fig:delta-tank-signal} shows a histogram of such quantity for both layers. In both layers, the histograms have a median of 0 PEs. The small asymmetrical tails (less than 1\% of the entries), which are due to tanks near the core, are a consequence of the limitation of the energy range of the LUT, which stops at 10 GeV. We did not expand further the LUT since the signal of those tanks is well above the saturation threshold of the simulated PMTs.

The difference in the number of entries is due to the lower layer having fewer tanks with at least 1 PE. We used for the histogram only tanks in which at least one of the two simulation algorithms predicted 1 PE. This was done to not bias the 0 PE bin with tanks in which no particle entered. Since the histograms in figure \ref{fig:delta-tank-signal} are extremely peaked around 0 and have long tails, the standard deviation is not representative of the dispersion of such quantities. To parametrize such a quantity, we have used the half width of the 84.1\% and 15.9\% quantiles of the distribution. For both histograms, we obtained less than 10 PEs.

\subsection{Tank time}
\begin{figure}
	\centering
	\includegraphics[width=0.6\linewidth]{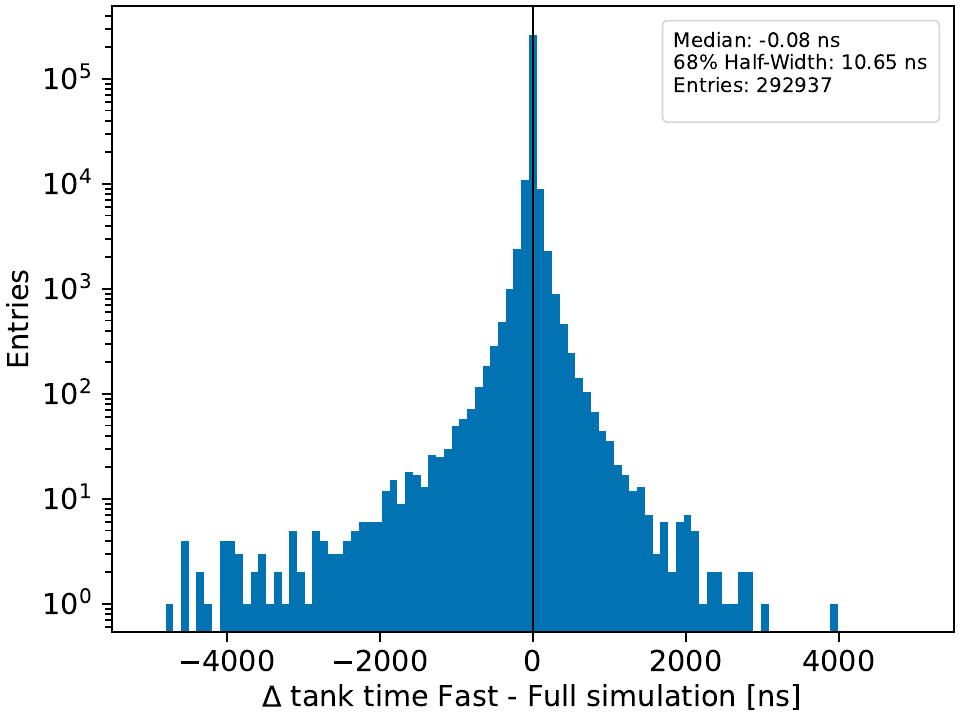}
	\caption{Difference of the arrival time of the signal in all tanks that detected at least 1 PE in both simulations. The logarithmic scale on the y-axis was used to show the very long tails of the distribution.}
	\label{fig:time-delta}
\end{figure}

It is possible to compute the difference in the first PE arrival time for all the tanks that detected some PEs in both simulations. Figure \ref{fig:time-delta} shows the histogram of such a variable. The histogram is symmetric and centered around 0, with tails that reach up to 1000 ns.

The distribution of the variable does not follow a gaussian profile, and computing the standard deviation would result in a not representative value. We have evaluated the position of the peak as the median of the distribution, which is -0.08 ns. For the dispersion, we have used the same procedure as for the difference of the tank signal. The value of such a quantity is 10.65 ns.

\section{Execution time reduction}
The authors evaluated the reduction of execution time with respect to the full simulation. We expect it to depend on the event core position since GEANT4 computing time should grow more than linearly with the number of particles crossing the detector, as the full simulation tracks all the Cherenkov photons inside the water. This analysis is long to perform, due to the need to run the full simulation code on UHE events, so we will show the results we obtained using another tank (in order not to run the test again). Since the layout we used in this evaluation is similar to the one used in this article, we do not expect relevant changes.

We produced 50 fixed-energy (100 TeV), fixed-direction (20$^\circ$) gamma showers and simulated them at fixed positions in the array with both fast and full simulations. Our initial idea was to start from the centre of the array and progress up to 500 m in steps of 50 m along the x-axis. However, the events which core was closer to the centre the array took more than 3 days, and the cluster put them on hold before they could end. Since fast simulation runs faster, we managed to finish the simulations within the time limit of the cluster.

\begin{figure}
	\centering
	\includegraphics[width=0.72\linewidth]{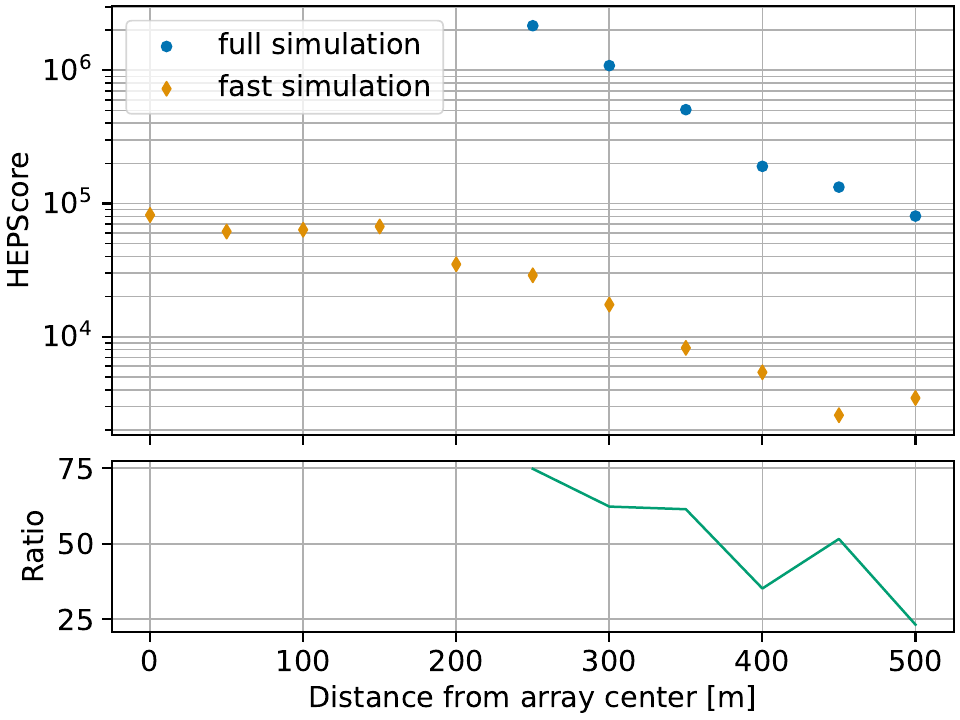}
	\caption{Top: HEPScore as a function of core position for both fast (orange diamonds) and full (cyan circles) simulation. Bottom: ratio of HEPScore between full and fast simulations.}
	\label{fig:time-gain}
\end{figure}

Figure \ref{fig:time-gain} shows the performance comparison as a function of core distance. Since we run the simulations on different machines (potentially with different performance), we have used the HEPScore \cite{hepscore} parameter as metric. We also computed the ratios of the HEPScores of full and fast simulation to get a reference value for the gain in computational time. Lastly, we double checked that using the normal CPU time of the various simulations did not completely change the results we saw with the HEPScore. We noticed that for the machines we used to run the test, one can obtain the CPU time by dividing by 10 the HEPScore.

For all simulated events, the gain in execution time is bigger than 20, surpassing 50 if we consider only events with the core inside the array. Our interpretation of this slope is that we reduce the computing time only when there are particles that cross the water (since we do not track them with the LUT approach). Therefore, for events really far away from the experiment, the gain with respect to the full simulation is smaller.

The trend followed by the ratio of the HEPScore confirms that the fast simulation is especially useful to simulate arrays with dense detector regions. Since the reduction of HEPScore is stronger near the center of the array, the fast simulation code improves gain in execution time with the number of particles that enter the detector. For this reason, as the number of particles in the EAS grows with the energy of the primary, we can also expect this reduction to improve with the energy of the primary.

\section{Conclusions}
We have developed an algorithm simulating the response of a detector array to extensive air showers without using a full detector simulation (based on GEANT4). The algorithm is based on a parametrisation of the detector response that has to be produced and tabulated in a look-up table. Furthermore, the algorithm only generates the response of the layout to EAS, it is still required to run CORSIKA to produce the input events.

The main aim of this work was to obtain a reliable simulation significantly faster than the standard GEANT4 approach. This requirement arises from the large event statistics required in high-energy studies (at least $10^5$ in each energy bin) and the time needed to simulate a single event in the UHE range (up to several hours). From our test array, we obtained a gain in time of at least a factor of 20, increasing for events well contained in the array, and so not falling outside.

The simulation was compared with the results of the full simulation. The average density of PEs shows a negligible bias (a few \%), but the fluctuations are up to 10\%. Overall, we found a systematic difference very near the core, mainly due to the discretization of the tank response (especially regarding the energy of the entering particle) and to the approximations done in the energy losses of particles crossing multiple detectors.

We never needed to utilize a complex interpolation of the parameters of the tank response, as we only used a nearest criteria. It could be interesting to explore the possibility of a more complicated code, to even more reduce the binning of the LUT or remove possible artifacts generated by the discreteness of the parametrisation. We also did not explore correlations within the tank response variables, assuming their independence.

Given the considerable reduction in computing time and the good agreement with the full simulation in the tested energy range and configurations, this alternative method can be a good compromise to reduce the computing requirements, proving a technical contribution that can support future studies for future cosmic or gamma ray experiments.
\section*{Acknowledgments}
\addcontentsline{toc}{section}{Acknowledgments}

We thank the SWGO and HAWC collaborations for providing the AERIE framework \cite{AERIE}, that we used to produce the simulations to build our example LUT and to perform comparison. We also acknowledge the support of the following agencies and organizations for the ongoing R\&D work towards SWGO: \url{https://www.swgo.org/SWGOWiki/doku.php?id=acknowledgements}.

\bibliographystyle{elsarticle-num} 
\bibliography{bibliography}

\end{document}